\documentclass[conference]{IEEEtran}
\usepackage[letterpaper,left=54pt,right=54pt,
bottom=54pt,top=72pt]{geometry} % First page
\IEEEoverridecommandlockouts
\usepackage{cite}
\usepackage{amsmath,amssymb,amsfonts}
\usepackage{algorithmic}
\usepackage{graphicx}
\usepackage{textcomp}
\usepackage{xcolor}
\usepackage{enumerate}
\usepackage{array,booktabs,arydshln,xcolor}
\usepackage[hidelinks]{hyperref}
\usepackage{caption}
\usepackage{comment}
\usepackage{siunitx}
\usepackage{stfloats}
\usepackage[colorinlistoftodos,textsize=scriptsize]{todonotes}

\newcommand{\cmmnt}[1]{}

\def\BibTeX{{\rm B\kern-.05em{\sc i\kern-.025em b}\kern-.08em
    T\kern-.1667em\lower.7ex\hbox{E}\kern-.125emX}}
    
\begin{document}

\newgeometry{letterpaper,left=54pt,right=54pt,
top=54pt,bottom=54pt}  % remaining pages

\title{On-Device Interpretable Tsetlin Machine-Based Intrusion Detection for Secure IoMT \\
}

\author{
    \IEEEauthorblockN{Rahul Jaiswal, Per-Arne Andersen, Linga Reddy Cenkeramaddi, Lei Jiao, Ole-Christoffer Granmo\\
    The Centre for Artificial Intelligence Research (CAIR) \\
    \IEEEauthorblockA{Department of ICT, University of Agder, Norway}
{\{rahul.jaiswal, per.andersen, linga.cenkeramaddi, lei.jiao, ole.granmo\}@uia.no}
}}

\begin{comment}
\IEEEoverridecommandlockouts \IEEEpubid{\makebox[\columnwidth]{979-8-3315-1276-8/26/\$31.00 \copyright 2026 IEEE \hfill} \hspace{\columnsep}\makebox[\columnwidth]{ }}
\end{comment}

\maketitle
%reviewcomment{The paper has a clear conference-shaped contribution, but it needs sharper positioning, stronger methodological traceability, and a cleaner submission build before submission.}

\begin{abstract} % 174 words % ok
The rapid evolution of digital health technologies is redefining healthcare services worldwide. The integration of wireless communication and Internet-enabled medical devices within Internet of Medical Things (IoMT) networks enables continuous, real-time patient monitoring. However, this increased connectivity raises cybersecurity and patient safety risks due to increasingly sophisticated cyberattacks. This paper proposes a novel on-device, interpretable Tsetlin Machine (TM)-based Intrusion Detection System (IDS) to identify various phases of cyberattacks in IoMT environments. The TM is a rule-driven and transparent machine learning (ML) approach that represents attack patterns using propositional logic. Extensive evaluations on the MedSec-25 dataset, encompassing various phases of realistic cyberattacks, show that the proposed model outperforms ML models, attaining an F1-score (macro) of 97.83\%. Moreover, the proposed model offers explicit explanations of its decisions to enhance transparency using feature-level contributions, class-wise vote scores, and clause activation heatmaps. Edge deployment (Raspberry Pi) further supports real-time on-device inference and intrusion detection. The combination of interpretability and high performance makes the proposed model well-suited for IoMT healthcare, where trust, reliability, safety, and timely decision-making are critical.
\end{abstract}
%\reviewcomment{The abstract should name the metric behind 97.83\% and avoid broad "outperforms state-of-the-art" language unless the comparison set is much broader than Table VI.}
%\reviewcomment{Also specify whether the reported 97.83\% is macro, micro, or weighted F1. For an imbalanced five-class IDS task, the averaging choice materially changes how the result should be interpreted.}


\section{Introduction} % ok
\label{intro}
% Internet of Medical Things (IoMT) and uses
The digital health technologies are advancing rapidly, enabling easier access to online doctor consultations through better internet connectivity, smartphones, and mobile applications. This approach is not only faster and more affordable for busy professionals but also highly convenient for elderly people, especially those living in rural or remote areas. It eliminates the need for unnecessary travel to the hospital, reduces waiting time, and offers a cost-effective alternative to the traditional in-person doctor visits.

\begin{figure}[t!]
\centering
\includegraphics[width=0.95\columnwidth,height=5.0cm,keepaspectratio]{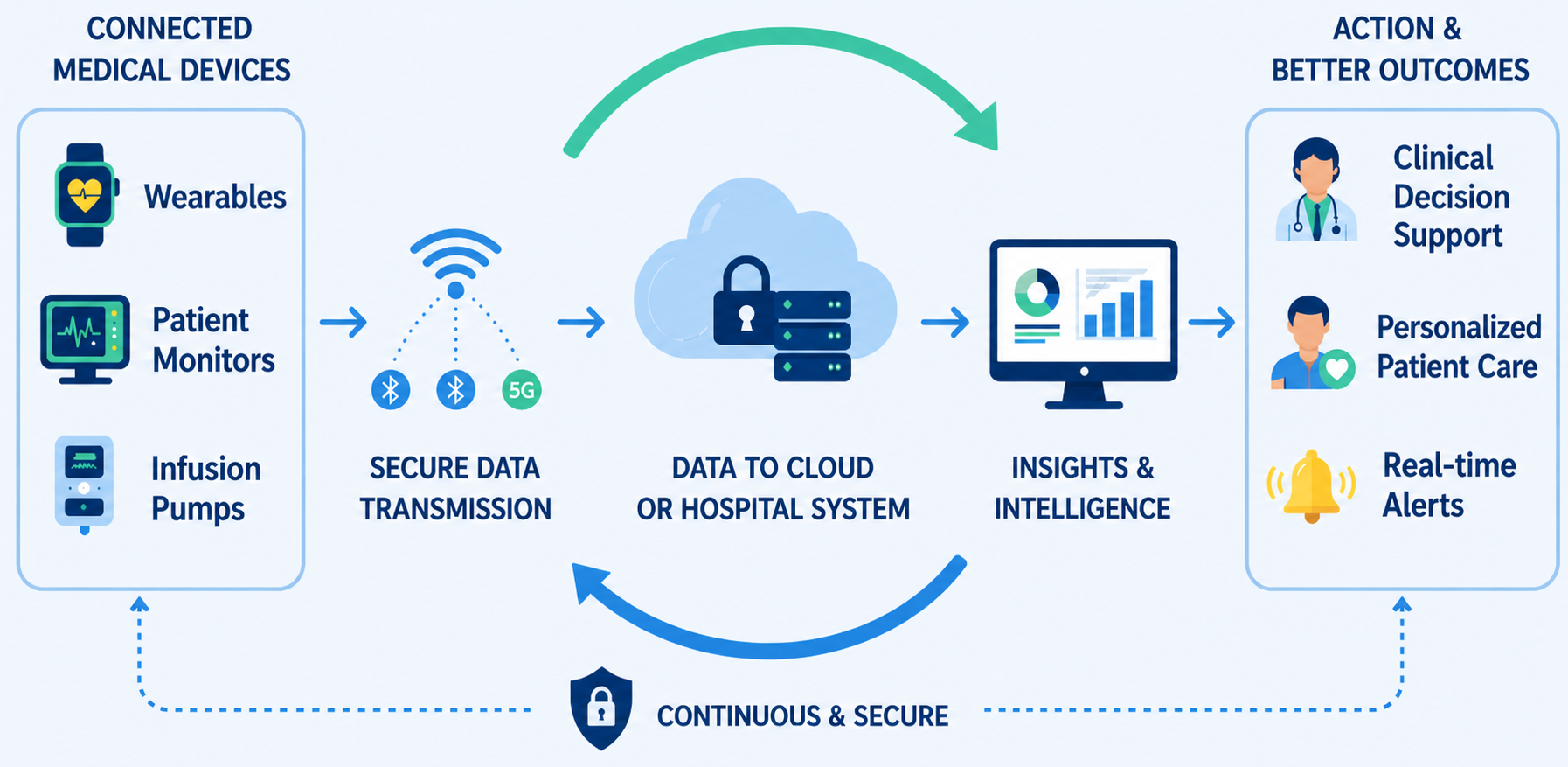} 
\caption{Overview of IoMT in healthcare.}
\label{iomt_app} \vspace{-3mm}
\end{figure}

In digital healthcare services, patients’ health data is transmitted electronically to the healthcare providers, such as hospitals and doctors, via digital channels using wearable medical devices (e.g., smart watches), as shown in Fig.~\ref{iomt_app}. The interconnected framework that enables medical devices, healthcare platforms, and digital systems to exchange medical data over the Internet is known as the Internet of Medical Things (IoMT)~\cite{huang2023internet}. It enables efficient transmission of health data from patient devices to healthcare providers, improving both the timeliness and effectiveness of ongoing patient care.

% Fig.1: IoMT working in healthcare.

% Challenges associated with the IoMT growth
The Deloitte Health Care Report 2026~\cite{report_deloitte} highlights that IoMT and digital health technologies are driving healthcare toward a ``Care Anywhere'' model, making virtual healthcare services more accessible and efficient. However, the use of IoMT medical devices introduces several critical challenges. Since these devices are connected to the Internet and continuously transmit highly sensitive medical data, they become attractive targets for cyberattacks. This raises major concerns such as unauthorized data access, medical device hijacking, exposure of sensitive personal health information, and cyberattacks that may disrupt system functionality or render medical devices unavailable, ultimately posing risks to both patient safety and privacy. For instance, if cyberattackers alter medical data of patients within hospital systems, then doctors may rely on the inaccurate readings, resulting in inappropriate treatment and potentially life-threatening harm to the patients. The CrowdStrike Global Threat Report 2026~\cite{report_crowdstrike} highlights that 10\% of global cyberattacks target healthcare systems.
%\reviewcomment{Verify both 2026 report claims before submission and give stable bibliographic metadata where possible. If these are web reports, include access dates or replace them with archival or DOI-backed sources.}

% Solution to IoMT cyberattacks
To ensure safe, secure, and reliable healthcare services, Intrusion Detection Systems (IDS)~\cite{diana2025overview} are used in the IoMT networks. They utilize network-based security techniques to monitor network traffic and identify potential cyberattacks in real time. Upon detecting suspicious activities, IDS promptly alerts the network administrator about potential intrusions or cyberattacks, enabling a timely response and mitigation. 

% About this paper work
This paper presents a novel IDS system for detecting different phases of cyberattacks targeting IoMT environments by leveraging an interpretable machine learning (ML) approach based on the Tsetlin Machine (TM)~\cite{kundu2026comprehensive, granmo2018tsetlin}. Building on this, network-level features are extracted from IoMT traffic and used as input to the TM model, which classifies the traffic as \texttt{benign (normal)} or \texttt{malicious attack phases}, thereby performing attack-phase classification. To promote transparency and build trust in the outcome of the TM model beyond black-box ML models, the interpretability of the TM model is explicitly analyzed. Moreover, the proposed TM-based IDS is deployed on the edge device (Raspberry Pi) to enable on-device inference for real-time intrusion detection.
%\reviewcomment{Differentiate the contribution more sharply from prior TM-based IDS work, especially the cited CICIoMT24 paper. State what is new here: MedSec-25, attack-phase classification, specific interpretability views, edge measurements, or all of these.}
%\reviewcomment{Define the prediction task consistently. The paper is not only classifying traffic as benign or malicious; the results use five classes. State clearly that this is attack-phase classification with benign plus four malicious phases.}

% Paper layout
The rest of this paper is structured as follows. Section~\ref{rel_work} reviews the related work and outlines the motivation. Section~\ref{bg} describes the classifiers employed and the on-device inference approach. Section~\ref{prop_ids} describes the proposed TM-based IDS. Section~\ref{exp_data} introduces the dataset. Section~\ref{res_dis} presents and discusses the results. Finally, Section~\ref{con_fut} concludes the paper and highlights directions for future research.

\section{Related Work and Motivation} % ok
\label{rel_work}
%\reviewcomment{Add a compact related-work comparison table: dataset, task type, number of classes, model family, interpretability support, edge deployment, and reported metric. This will make the research gap clearer than paragraph-by-paragraph enumeration.}
% Literature review on IDS
% Traditional IDS with demerits
Intrusion Detection Systems (IDS) play a critical role in safeguarding sensitive medical data from cyberattacks within IoMT environments. It can detect both internal and external attacks that bypass existing security measures. Traditional IDS systems are broadly categorized into two approaches, namely, signature-based and anomaly-based detection~\cite{nawaal2024signature,bhavsar2023anomaly}. The signature-based approach identifies known attacks by matching observed patterns against predefined signatures. In contrast, anomaly-based detection focuses on identifying deviations from the normal network behavior to detect potential attacks. However, the highly dynamic nature of IoMT environments limits the effectiveness of both approaches at identifying new or unseen cyberattacks, leaving IoMT devices exposed to emerging intrusions.

% ML and DL based IDS with demerits
Machine learning (ML) and deep learning (DL) based intrusion detection systems have demonstrated strong potential for enhancing IoMT security. %These techniques can identify attacks by detecting deviations in the attributes of network traffic. 
For instance, the work in~\cite{anitha2023artificial} investigates various ML approaches for cyberattack detection on the IEEE DataPort dataset and attains an accuracy of 89.89\% for binary classification using K-nearest Neighbours (KNN). In~\cite{awotunde2021deep}, a deep autoencoder-based IDS is developed to secure IoMT using the NF-ToN-IoT dataset, achieving an accuracy of 89\% for 10-class classification. In~\cite{kavkas2025enhancing}, the authors propose deep neural network (DNN) and long short-term memory (LSTM) based models for cyberattack detection in IoMT environments using the CICIoMT24 dataset, achieving accuracies of 78\% and 79\%, respectively, for six-class classification. In~\cite{dadkhah2024ciciomt2024}, the authors apply various ML techniques for cyberattack detection, achieving an accuracy of 73.5\% in a six-class classification setting. Although these ML and DL models can attain high classification accuracy, interpretability remains a critical requirement for their effective adoption in healthcare services. Moreover, many of these models operate as black boxes, offering limited insight into their decision-making processes, raising significant concerns about the transparency of their outcomes. The deployment of these models on edge devices has also not been investigated. %The Tsetlin Machine (TM)~\cite{kundu2026comprehensive, granmo2018tsetlin} has emerged as a promising alternative to address these limitations.

% TM-based IDS with benefits
The TM is an emerging ML technique based on Tsetlin Automata~\cite{granmo2018tsetlin}, well-suited for IoMT intrusion detection. It focuses on identifying recurring patterns in network traffic through frequent pattern learning and efficient resource allocation, rather than relying solely on error minimization. By representing knowledge as interpretable conjunctive clauses, the TM reduces overfitting and simplifies intrusion detection into manageable patterns, making it suitable for reliable intrusion detection. For instance, the study in~\cite{gunvaldsen2023towards} introduces a TM-based anomaly detection framework and evaluates it across multiple datasets to show improved accuracy than ML classifiers. TM is integrated with transfer learning~\cite{jaiswal2025leveraging, jaiswal2023location, jaiswal2025data} to classify cervical cancer using the
pap smear image dataset in~\cite{ahishakiye2026enhanced}, highlighting TM's adaptability and interpretable rule-based learning, which are beneficial for intrusion detection. A TM-based IDS using the CICIoMT24 dataset is developed in~\cite{jaiswal2026tsetlin}, achieving superior accuracy than ML and DL classifiers. However, feature-level interpretability and edge-device deployment remain unexplored. These findings motivate us to propose a novel, transparent, and interpretable TM-based IDS for attack-phase detection in IoMT environments and demonstrate its deployment on an edge device such as a Raspberry Pi. This paper makes the following contributions:
\begin{itemize}
    \item Design and implementation of an effective TM-based IDS for attack-phase detection in IoMT environments using the MedSec-25 dataset.
    %\item Evaluation of the proposed IDS in a multi-phase attack detection setting using the MedSec-25 dataset.
    \item Numerical evaluation demonstrating the superiority of the proposed IDS over various ML methods.
    \item Interpretability analysis of the proposed IDS to reveal decision rules for accurate attack-phase detection.
    \item Edge deployment of the proposed IDS on a Raspberry Pi for real-time on-device inference.
    %\item Demonstration of the practical applicability of the TM-based solutions for IoMT security. 
\end{itemize}
%\reviewcomment{The contribution claims overlap. Consider compressing them into 3-4 distinct points: model/design, evaluation protocol, interpretability, and edge deployment.}
%\reviewcomment{The cervical-cancer TM example feels tangential in this short IEEE paper. Either explain why it matters for the transferability of TM interpretability, or replace it with more directly relevant IDS/TM security work.}

\section{Background} % ok
\label{bg}
In this section, Tsetlin Machines, classification models, and an on-device inference approach are introduced.

\subsection{Tsetlin Machine}
\label{tmc}
%\reviewcomment{The claim that TM is robust to class imbalance needs evidence or qualification. If this is based on prior work, cite it; if it is based on these experiments, show class-wise recall and compare against imbalance-handling alternatives.}
The TM is an interpretable, rule-based machine learning model that constructs logical clauses using propositional logic~\cite{granmo2018tsetlin}. It represents cyberattack patterns as human-readable logical expressions, thereby enabling transparent and explainable intrusion detection. Owing to its binary feature representation, TM is particularly suitable for deployment on resource-constrained devices. %Additionally, TM shows strong robustness to class imbalance, a prevalent feature of IoMT datasets such as MedSec-25 (see Table~\ref{summary_attack}).

The TM comprises a set of conjunctive clauses defined over binary input features. A clause $C_j$ is expressed as~\cite{granmo2018tsetlin}:
\begin{equation}
C_j = \bigwedge_{k \in I_j} x_k \ \wedge \ \bigwedge_{l \in \bar{I}_j} \neg x_l
\label{eq:clause}
\end{equation}
where $x_k \in \{0,1\}$ denotes binary features, and $I_j$ and $\bar{I}_j$ are the sets of included and negated literals, respectively.

Each clause casts a vote toward or against a class, and the final prediction is obtained by aggregating clause output as:
\begin{equation}
f(\mathbf{x}) = \sum_{j=1}^{m} w_j C_j(\mathbf{x}), \quad \text{with} \quad |f(\mathbf{x})| \leq T
\label{eq:voting}
\end{equation}
where $w_j \in \{+1, -1\}$ denotes the polarity of clause $j$, and $T$ is the voting threshold that constrains the total clause contribution, ensuring balanced learning. The clause granularity is controlled by the specificity parameter $s > 1$, which regulates the inclusion probability of literals. 

\textit{Sparse Tsetlin Machine (STM)}: It is a variant of the TM in which, unlike the standard TM that considers all literals (both features and their negations) when forming clauses, only a subset of literals is used per clause. The STM introduces controlled sparsity, leading to fewer active features, reduced influence of irrelevant features, and improved generalization. A clause $C_j$ in STM is expressed as~\cite{granmo2018tsetlin}:
\begin{equation}
C_j = \bigwedge_{k \in S_j^{+}} x_k \ \wedge \ \bigwedge_{l \in S_j^{-}} \neg x_l
\label{eq:sparse_clause}
\end{equation}
where $S_j^{+} \subseteq \{1,\dots,n\}$ and $S_j^{-} \subseteq \{1,\dots,n\}$ denote the subsets of selected features included as positive and negated literals in clause $j$, respectively, with $|S_j^{+} \cup S_j^{-}| \ll n$. $n$ denotes the number of input features (see Table~\ref{para_TM_ML}).
%\reviewcomment{Check the set notation: if braces should appear, use \texttt{\textbackslash\{1,\textbackslash dots,n\textbackslash\}}. More importantly, connect this STM description to how STM is configured in the experiments.}

% Multi-class classification using TM
In multi-class classification, each class is represented by a dedicated set of clauses, which are evenly divided into positive and negative polarity groups. These clauses are formed using combinations of input features and their negations to capture class-specific patterns. For a given input, each clause evaluates to either true or false, and contributes a vote toward or against its associated class. The overall score for each class is obtained by aggregating the votes from its positive and negative clauses. Finally, the predicted label is determined by selecting the class with the highest aggregated score. The detailed mathematical formulation of TM can be seen in~\cite{jaiswal2026tsetlin}.

% Fig.2: Proposed TM-based IDS.
\begin{figure*}[b!]
\centering
\includegraphics[width=\textwidth,height=8.5cm,keepaspectratio]{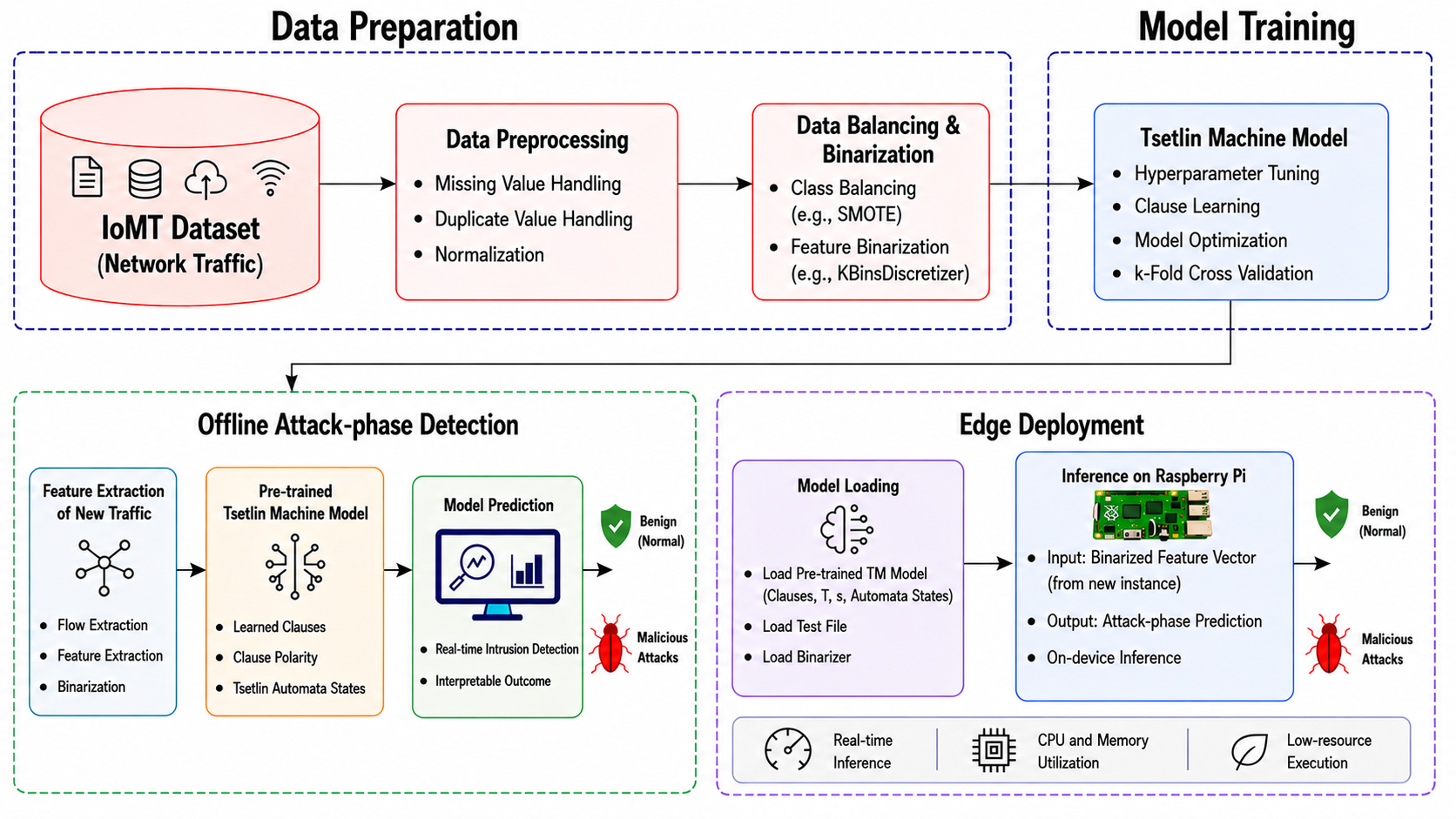} 
\caption{Proposed TM-Based IDS architecture.}
\label{ids_tm} %\vspace{-3mm}
\end{figure*}

\subsection{Classification Models} % ok
\label{mlc}

\subsubsection{Decision Tree (DT)}
DT is a tree-based classifier that recursively partitions feature space using decision rules~\cite{breiman2017classification}.

\subsubsection{Random Forest (RF)}
RF is an ensemble of decision trees that determines final class through majority voting~\cite{jedari2015wi}.

\subsubsection{XGBoost}
XGBoost is a gradient-boosting framework that sequentially constructs decision trees to improve predictive performance~\cite{chen2016xgboost}.

\subsubsection{LightGBM (LGBM)}
LightGBM is an efficient gradient-boosting framework that employs histogram-based learning and leaf-wise tree growth~\cite{ke2017lightgbm}.

\subsubsection{K-Nearest Neighbours (KNN)}
It classifies a sample according to majority class among its nearest neighbours~\cite{jedari2015wi}.

\subsubsection{Naive Bayes (NB)}
NB is a probabilistic classifier based on Bayes' theorem and the conditional independence assumption~\cite{alpaydin2020introduction,jaiswal2023caqoe}.

\subsubsection{Logistic Regression (LR)}
LR is a linear classifier that estimates class probabilities using a logistic function~\cite{jaiswal2023caqoe}.

\subsubsection{Neural Network (NN)} 
NN is a multi-layer learning model that learns complex nonlinear patterns from data~\cite{jaiswal2023caqoe}.

\subsection{On-Device Inference} % ok
\label{ed}
%\reviewcomment{The Raspberry Pi hardware description can be shortened. Use the saved space to report the measurement procedure: batch size, number of repetitions, warm-up, CPU sampling method, and whether preprocessing or feature extraction was timed.}
% About Raspberry Pi 5
To evaluate on-device inference of the TM model on a resource-constrained platform, Raspberry Pi 5 Model B~\cite{mathe2024comprehensive} is employed. It is a compact and efficient computing platform, approximately the size of a credit/debit card, and serves as a powerful mini-computer (see Fig.~\ref{ids_tm}). It has a quad-core ARM Cortex-A76 processor operating at 2.4~GHz and supports up to 8~GB of SDRAM. The board includes two USB 3.0 and two USB 2.0 ports, and supports Ethernet, Wi-Fi, and Bluetooth connectivity at 2.4~GHz and 5~GHz. Its on-device deployment capability facilitates efficient execution of the TM model in the resource-constrained IoMT environments.

% Experimental setup of Raspberry Pi 
The Raspberry Pi is connected to the host computer via Ethernet, enabling reliable network communication between the devices. The TM and other ML models are implemented in Python on Raspberry Pi for real-time on-device inference. %, enabling on-device inference, while development and data transfer are managed from the host system. This configuration supports seamless deployment and real-time on-device inference on the edge device, that is, the Raspberry Pi.

\section{Proposed Intrusion Detection System} % ok
\label{prop_ids}
%\reviewcomment{Give the exact preprocessing order in this section. For example: split first, remove leakage-prone columns, fit scaler on training data, fit binarizer on training data, apply SMOTE to training data only, then train. The order is central to reproducibility.}
The objective of the proposed IDS is to accurately detect attack phases in IoMT environments. The overall architecture of the proposed TM-based IDS is illustrated in Fig.~\ref{ids_tm}, comprising various stages, namely, data preparation, model training and testing, and edge deployment.

In the first stage, data preparation is carried out to construct a suitable input for training the TM model. This stage involves selecting the IoMT dataset, followed by preprocessing steps such as handling missing and duplicate entries, normalizing the data, and addressing class imbalance. Subsequently, feature binarization is performed to transform the input data into a binary format compatible with the TM model.

In the second stage, the TM model is trained to learn discriminative patterns from the processed data. This includes hyperparameter tuning, clause formation, and model optimization. For reliable performance and to reduce overfitting, k-fold cross-validation~\cite{bhagwat2019applied} is employed. During training, the TM learns interpretable logical relationships in the form of positive and negative clauses to compute class-specific scores for classification. Using these learned clauses, the pre-trained TM model classifies a given test network traffic instance as either \texttt{benign} or \texttt{malicious attack phases}. Next, the trained model, along with its learned clauses and automata states, is then stored for deployment.

In the deployment stage, the trained model is deployed on an edge device using the Raspberry Pi to enable real-time on-device inference. The pre-trained TM model, binarizer, and test file are loaded onto the device, after which the model performs on-device classification to label the traffic as either \texttt{benign} or \texttt{malicious attack phases}. This edge-based implementation enables low-latency detection and reduces reliance on the centralized systems.

\begin{table*} [b!]
\caption{Feature set of the MedSec-25 dataset.}
\label{summary_feature} 
\centering
{
\setlength\tabcolsep{0.5pt}
\begin{tabular}{|c|c|c|c|c|c|c|c|c|c|}
\hline
S.No. & Feature name & S.No. & Feature name & S.No. & Feature name & S.No. & Feature name & S.No. & Feature name \\
\hline
1. & \texttt{Flow ID} & 2. & \texttt{Src IP} & 3. & \texttt{Src Port} & 4. & \texttt{Dst IP} & 5. & \texttt{Dst Port} \\ 
6. & \texttt{Protocol} & 7. & \texttt{Timestamp} & 8. & \texttt{Flow Duration} & 9. & \texttt{Tot Fwd Pkts} & 10. & \texttt{Tot Bwd Pkts} \\  
11. & \texttt{TotLen Fwd Pkts} & 12. & \texttt{TotLen Bwd Pkts} & 13. & \texttt{Fwd Pkt Len Max} & 14. & \texttt{Fwd Pkt Len Min} & 15. & \texttt{Fwd Pkt Len Mean} \\ 
16. & \texttt{Fwd Pkt Len Std} & 17. & \texttt{Bwd Pkt Len Max} & 18. & \texttt{Bwd Pkt Len Min} & 19. & \texttt{Bwd Pkt Len Mean} & 20. & \texttt{Bwd Pkt Len Std} \\
21. & \texttt{Flow Byts/s} & 22. & \texttt{Flow Pkts/s} & 23. & \texttt{Flow IAT Mean} & 24. & \texttt{Flow IAT Std} & 
25. & \texttt{Flow IAT Max} \\ 
26. & \texttt{Flow IAT Min} & 27. & \texttt{Fwd IAT Tot} & 28. & \texttt{Fwd IAT Mean} & 29. & \texttt{Fwd IAT Std} & 30. & \texttt{Fwd IAT Max} \\
31. & \texttt{Fwd IAT Min} & 32. & \texttt{Bwd IAT Tot} & 33. & \texttt{Bwd IAT Mean} & 34. & \texttt{Bwd IAT Std} & 35. & \texttt{Bwd IAT Max} \\ 
36. & \texttt{Bwd IAT Min} & 37. & \texttt{Fwd PSH Flags} & 38. & \texttt{Bwd PSH Flags} & 39. & \texttt{Fwd URG Flags} & 40. & \texttt{Bwd URG Flags} \\ 
41. & \texttt{Fwd Header Len} & 42. & \texttt{Bwd Header Len} & 43. & \texttt{Fwd Pkts/s} & 44. & \texttt{Bwd Pkts/s} & 45. & \texttt{Pkt Len Min} \\
46. & \texttt{Pkt Len Max} & 47. & \texttt{Pkt Len Mean} & 48. & \texttt{Pkt Len Std} & 49. & \texttt{Pkt Len Var} & 50. & \texttt{FIN Flag Cnt} \\ 
51. & \texttt{SYN Flag Cnt} & 52. & \texttt{RST Flag Cnt} & 53. & \texttt{PSH Flag Cnt} & 54. & \texttt{ACK Flag Cnt} & 55. & \texttt{URG Flag Cnt} \\
56. & \texttt{CWE Flag Count} & 57. & \texttt{ECE Flag Cnt} & 58. & \texttt{Down/Up Ratio} & 59. & \texttt{Pkt Size Avg} & 60. & \texttt{Fwd Seg Size Avg} \\
61. & \texttt{Bwd Seg Size Avg} & 62. & \texttt{Fwd Byts/b Avg} & 63. & \texttt{Fwd Pkts/b Avg} & 64. & \texttt{Fwd Blk Rate Avg} & 65. & \texttt{Bwd Byts/b Avg} \\
66. & \texttt{Bwd Pkts/b Avg} & 67. & \texttt{Bwd Blk Rate Avg}  & 68. & \texttt{Subflow Fwd Pkts} & 69. & \texttt{Subflow Fwd Byts} & 70. & \texttt{Subflow Bwd Pkts} \\
71. & \texttt{Subflow Bwd Byts} & 72. & \texttt{Init Fwd Win Byts} & 73. & \texttt{Init Bwd Win Byts} & 74. & \texttt{Fwd Act Data Pkts} & 75. & \texttt{Fwd Seg Size Min} \\
76. & \texttt{Active Mean} & 77. & \texttt{Active Std} & 78. & \texttt{Active Max} & 79. & \texttt{Active Min} & 80. & \texttt{Idle Mean} \\
\cline{9-10}
81. & \texttt{Idle Std} & 82. & \texttt{Idle Max} & 83. & \texttt{Idle Min} & 84. & \texttt{Label} & \multicolumn{2}{c|}{Total features = 84}  \\
\hline
\end{tabular} \vspace{-3mm}
}
\end{table*}

% Table 2- MedSec-25 attack phases and number of samples.
\begin{table} [t!]
\caption{Attack phases in the MedSec-25 dataset.}
\label{summary_attack} 
\centering
{
\setlength\tabcolsep{10.0pt}
\begin{tabular}{|c|c|}
\hline
Attack phases & Sample count \\
\hline
Benign & 12,348 \\
\hline
Reconnaissance & 401,683 \\
\hline
Initial access & 102,090 \\
\hline
Lateral movement & 12,498 \\
\hline
Exfiltration & 25,915 \\
\hline
%\multicolumn{2}{|c|}{Total sample count = 554,534} \\
%\hline
\end{tabular} \vspace{-4mm}
}
\end{table}

\section{Experimental Dataset} % ok
\label{exp_data}
To enhance IoMT security, the MedSec-25 dataset\footnote{Link to dataset: https://www.kaggle.com/datasets/abdullah001234/medsec-25-iomt-cybersecurity-dataset (accessed on February 10, 2026).}~\cite{almobaideen2025medsec} presents the behavior of benign and malicious traffic within a realistic healthcare IoT network environment that reflects real-world hospital operations. The setup utilizes Raspberry Pi nodes along with various medical and environmental sensors, including Electrocardiogram (ECG), Electroencephalogram (EEG), respiration, thermistor, ultrasonic modules, and environmental sensors. It employs a variety of IoMT devices and protocols, including Message Queuing Telemetry Transport (MQTT), Secure Shell (SSH), File Transfer Protocol (FTP), Hypertext Transfer Protocol (HTTP), and Domain Name System (DNS). The benign traffic is generated using multiple services and operations, such as HTTP-based services and real user interactions with medical devices. A multi-stage attack campaign is conducted to simulate malicious traffic, which facilitates early detection of attacks at each phase and limits further escalation. Based on the MITRE ATT\&CK framework~\cite{att2024mitre}, the attack campaign comprises four phases, namely, reconnaissance, initial access, lateral movement, and exfiltration, capturing the major phases of real-world attacks. The generated network traffic is captured using Wireshark and stored in PCAPNG format. The flow-based network features are extracted using CICFlowMeter, and CSV files are generated for each attack phase. 

% About different attack phases
Reconnaissance is the initial phase of cyberattacks, in which an adversary/cyberattacker gathers information about the target network, including active hosts, open ports, and system characteristics, using techniques such as SYN/TCP scanning and operating system (OS) fingerprinting. During the initial access phase, the adversary gains unauthorized entry into a target system or network by exploiting vulnerabilities, weak credentials, or exposed services (e.g., SSH, FTP, or web applications). After gaining initial access, the lateral movement phase enables the adversary to propagate across the network and compromise additional systems by leveraging credentials, remote services, or network weaknesses. Finally, in the exfiltration phase, the cyberattacker extracts sensitive data from compromised systems and sends it to external destinations using covert communication channels. The dataset is imbalanced, comprising 84 features and 554,534 samples, as presented in Table~\ref{summary_feature} and Table~\ref{summary_attack}, respectively. 

\section{Results and Discussions} % ok
\label{res_dis}
This section describes the experimental setup and evaluation methodology, followed by an in-depth analysis of results.

\subsection{Experimental Setup} % ok
\label{sys_set}
All algorithms are implemented in Python~3.13.6 using the Keras framework built on TensorFlow~2.20.0. NumPy~2.3.2, Pandas~2.3.1, scikit-learn~1.7.2, imbalanced-learn~0.14.0, XGBoost~3.2.0, and LightGBM~4.6.0 are used for data processing and model development. The experiments are conducted on a MacBook equipped with an Apple M4 chip and 16~GB of RAM. The on-device inference experiments are conducted on a Raspberry Pi~5 using Python~3.11.9 and TMU~0.6.5.
%\reviewcomment{Verify the Python/TensorFlow version combination. TensorFlow 2.2 is very old relative to Python 3.13, so this line may be read as an environment-reporting error unless corrected or explained. Add versions for scikit-learn, imbalanced-learn, xgboost, lightgbm, numpy, pandas, and TMU.}

%\vspace{-1mm}
% Performance Evaluation
\subsection{Evaluation Methodology} % ok
\label{eva_method} 
To evaluate classification performance, the models are assessed using macro-averaged precision, recall, and F1-score, which treat all classes equally and are suitable for imbalanced datasets. Precision reflects the correctness of detected attacks, while recall measures the capability of the IDS to identify actual attacks. The F1-score combines these two metrics into a single measure by balancing precision and recall. In multi-class classification, the F1-score is generally preferred over accuracy, as it accounts for both false positives and false negatives. This is particularly important for imbalanced datasets (e.g., in our case of the MedSec-25 dataset), where accuracy can be dominated by majority classes and may not accurately represent performance across all classes. For instance, failing to detect rare attack phases (low recall) or incorrectly classifying benign traffic as malicious attack phases (low precision) can significantly impact system reliability. Such trade-offs are not adequately captured by accuracy, whereas the F1-score provides a more comprehensive evaluation. The class-wise precision, recall and F1-score are also presented for the proposed TM model. To this end, the dataset was partitioned into 80\% training and 20\% testing subsets using stratified random sampling with a fixed random seed of 42. The training data are balanced using SMOTE, normalized, and binarized before applying five-fold stratified cross-validation~\cite{bhagwat2019applied}. The performance metrics are averaged across the validation folds, with empirically fixed hyperparameter settings throughout cross-validation. A confusion matrix is employed to demonstrate the classification results and analyze class-wise performance. The proposed IDS is evaluated against ML classifiers (Section~\ref{mlc}) and state-of-the-art approaches. Moreover, inference time, defined as the duration required by a trained model to produce the prediction for a given input sample during deployment, is measured and compared to assess computational efficiency.

To showcase the interpretability of the proposed TM model, class-specific vote scores and clause activation heatmaps are analyzed to explain the classification decisions. A higher vote score for a given class reflects stronger evidence that the input network traffic belongs to that class, thereby indicating a more confident classification decision. In addition, feature-level contributions are examined to identify the key features which influence correct predictions. The performance metrics are defined as~\cite{jaiswal2022performance}:
\begin{equation}
\text{Precision} = \frac{TP}{TP + FP},~~\text{Recall} = \frac{TP}{TP + FN},
\end{equation}

\begin{equation}
\text{F1-score} = \frac{2 \times (\text{Precision} \times \text{Recall})}
{\text{Precision} + \text{Recall}},
\end{equation}
where $TP$, $TN$, $FP$, and $FN$ denote the true positive, true negative, false positive, and false negative, respectively. For the multi-class attack-phase classification task, these quantities are computed using a one-vs-rest convention, where each class is treated as the positive class, and all remaining classes are treated as the negative class.
%\reviewcomment{For a five-class problem, define TP, FP, TN, and FN per class or state that a one-vs-rest convention is used. The current wording reads like a binary IDS formulation.}

% Fig.3: Data imbalance 
\begin{figure}[b!]
\centering
\includegraphics[width=0.9\columnwidth,height=3.9cm,keepaspectratio]{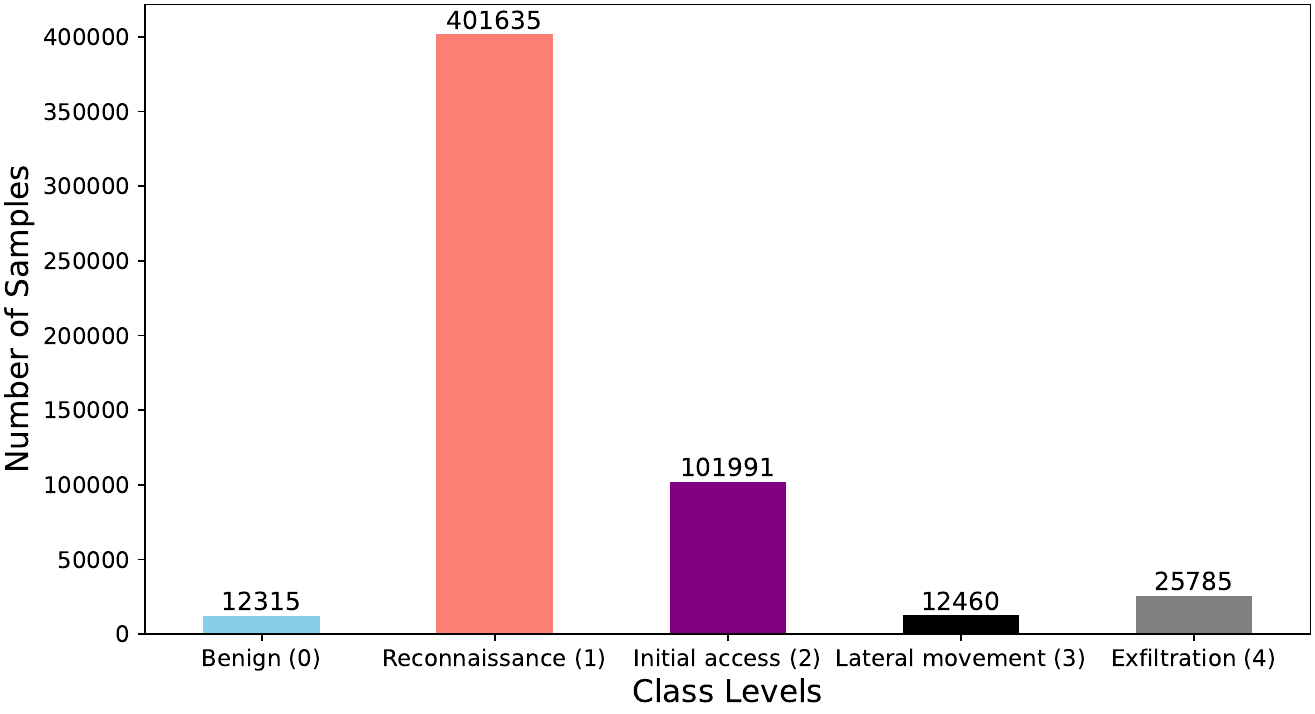} 
\caption{Class imbalance: Multi-class (five-class) classification.}
\label{clas_imb} \vspace{-3mm}
\end{figure}

% Fig.4: SMOTE used for data imbalance
\begin{figure}[b!]
\centering
\includegraphics[width=0.9\columnwidth,height=3.9cm,keepaspectratio]{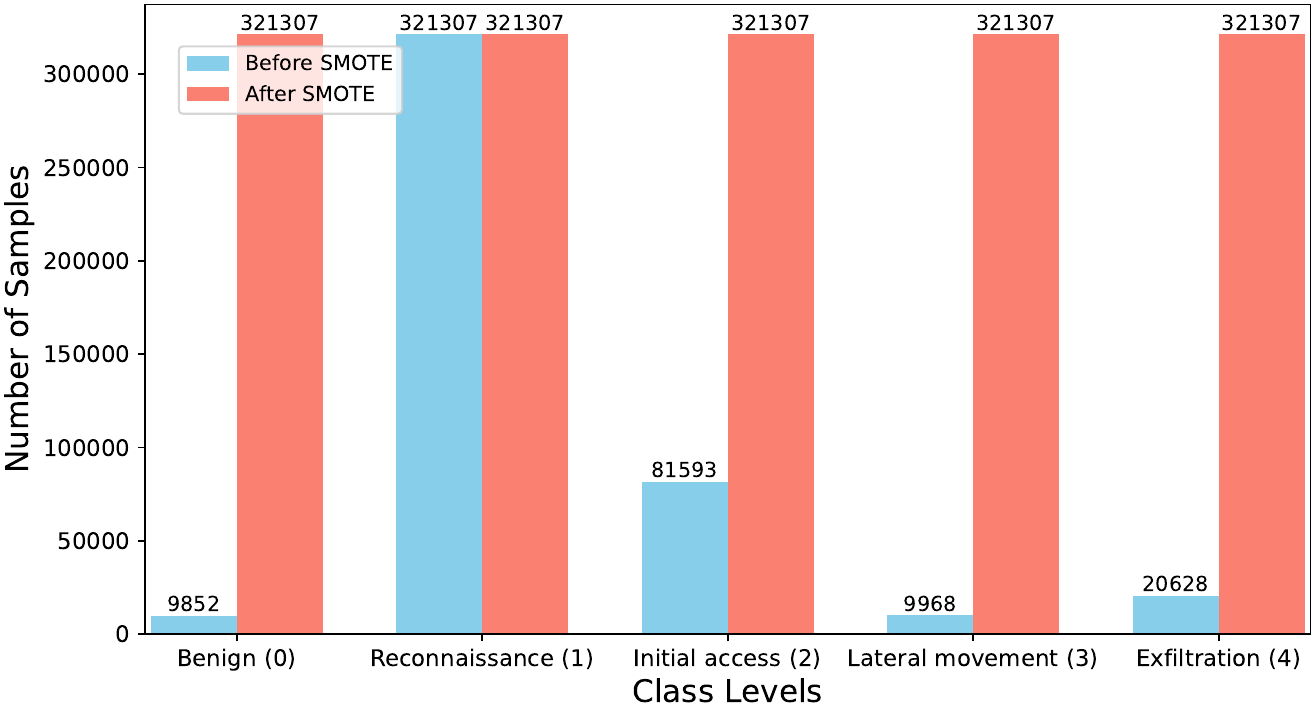} 
\caption{Balanced training classes.}
\label{clas_imb_smote} \vspace{-3mm}
\end{figure}

\subsection{Classification Analysis} % ok
\label{clas_ana}
\subsubsection{Data Pre-processing} The dataset (see Table~\ref{summary_attack}) exhibits class imbalance and contains a few duplicate samples (348). To enhance data quality, these samples are eliminated. The categorical attack labels are numerically encoded, with 0 representing benign traffic and 1--4 representing attack phases of Reconnaissance, Initial Access, Lateral Movement, and Exfiltration. The Flow ID, Src IP, Dst IP, Timestamp, and Label features are dropped to eliminate identifier-based bias and potential data leakage. Despite this cleaning process, the dataset remains imbalanced, as depicted in Fig.~\ref{clas_imb}. The dataset is then partitioned using a stratified random 80/20 train-test split with a fixed random seed of 42. Training models on such imbalanced data can lead to degraded performance and a higher rate of false positives. To mitigate this issue, different strategies are adopted for the TM and ML models. For the TM model, the Synthetic Minority Over-sampling Technique (SMOTE)~\cite{bhagwat2019applied} is applied exclusively to the training set to generate synthetic samples for minority classes, as depicted in Fig.~\ref{clas_imb_smote}. In contrast, for the ML classifiers, class imbalance is handled using the \text{compute\_class\_weight} method~\cite{bhagwat2019applied}, which assigns higher weights to minority classes and lower weights to majority classes during training. These strategies reflect commonly adopted imbalance-handling practices for the respective model families and help in reducing bias toward the majority classes. %Notably, SMOTE is not applied to the test set, thereby preserving the original data distribution and ensuring an unbiased evaluation
%\reviewcomment{Add the dataset version or date, exact files used, and whether the split was flow-random, time-aware, device-aware, or scenario-aware. This matters for leakage risk in network traffic datasets.}
%\reviewcomment{The class imbalance is severe, especially for benign and lateral movement relative to reconnaissance. This should motivate macro-F1 and per-class recall in the main results, not only aggregate precision, recall, and F1.}
%\reviewcomment{Report how many rows were removed as missing or duplicate, and whether duplicate detection was performed before or after splitting. Removing duplicates globally can be acceptable, but the handling should be transparent.}
%\reviewcomment{Clarify that this table includes non-model columns such as label and likely identifiers. Then state exactly which columns were excluded from model input; the NN table later suggests 79 input features, but this table lists 84 including Label.}
%\reviewcomment{If five-fold cross-validation is used, SMOTE and discretization should be fitted inside each training fold only. State this explicitly to avoid concerns about leakage.}
%\reviewcomment{Using SMOTE for TM but class weights for the ML baselines may make the comparison method-dependent. Either apply comparable imbalance strategies across model families, or justify why each model receives a different treatment.}

\subsubsection{Classifier Training} 
 The numerical features are first normalized using min-max scaling, fitted on the training data and applied to the test data. This step improves training stability and ensures that each feature contributes equally to the learning process. As the TM operates on binary inputs and relies on logical rule formation, these normalized features are further transformed into discrete intervals using the KBinsDiscretizer~\cite{bhagwat2019applied}, facilitating effective binarization and interpretable clause learning. 
Similarly, the same preprocessed data (without binarization) is used to train various ML classifiers (see Section~\ref{mlc}). The parameters for all models are empirically determined, with a random seed of 42 used throughout. The parameter settings and classification performance of the TM and ML models are summarized in Table~\ref{para_TM_ML} and Table~\ref{mod_perf}, respectively. %Additionally, Fig.~\ref{train_test_acc_TM} shows the training and testing accuracy for the TM model, illustrating stable convergence during training.
%\reviewcomment{Add the tuning procedure and random seeds so the comparison is reproducible.}
%\reviewcomment{State whether normalization was fitted only on the training data. If the scaler was fitted before the split or before CV folds, the reported performance can be optimistically biased.}

% Table 3- Parameters used in TM and ML classifiers
\begin{table} [t!]
\caption{Model parameters.}
\label{para_TM_ML} 
\centering
{
\setlength\tabcolsep{3.0pt}
\begin{tabular}{|c|c|}
\hline
Model & Parameters \\
\hline
TM & Binarizer: KBinsDiscretizer, \text{n\_bins}=40, \\ 
& encode=\text{onehot-dense}, strategy=quantile \\
\cline{2-2}
& \text{number\_of\_clauses}=2000, $T$=30, $s$=15, \\
& \text{weighted\_clauses}=False, Epochs=45 \\
\hline
STM & Binarizer: KBinsDiscretizer, \text{n\_bins}=25, \\ 
& encode=\text{onehot-dense}, strategy=quantile \\
\cline{2-2}
& \text{number\_of\_clauses}=1500, $T$=30, $s$=20, \\
& \text{weighted\_clauses}=False, \text{literal\_sampling}=1 \\ 
& \text{max\_included\_literals=16}, absorbing=0, Epochs=50 \\
\hline
DT, RF & \text{compute\_class\_weight} method with \text{class\_weight=balanced} \\
\hline
XGBoost & objective=multi:softprob, \text{eval\_metric}=mlogloss, \\
& \text{tree\_method}=hist, \text{learning\_rate}=0.1, \\ & \text{max\_depth}=8, \text{n\_estimators=200} \\
\hline
LGBM & objective=multiclass, \text{learning\_rate}=0.1, \text{n\_estimators=200}, \\
& \text{compute\_class\_weight} method with \text{class\_weight=balanced} \\
\hline
KNN & \text{n\_neighbors=5} \\
\hline
NB & default settings \\
\hline
LR & solver=lbfgs, max\_iter=500, multi\_class=multinomial, \\
& \text{compute\_class\_weight} method with \text{class\_weight=balanced} \\
\hline
NN & input layer=79 neurons, first hidden layer=64 neurons, \\
& second hidden layer=32 neurons, output layer=1 neurons, \\
& hidden layer activation function=relu, \\
& output layer activation function=softmax, \\
& optimizer=adam, \text{loss=sparse\_categorical\_crossentropy}, \\
& metrics=accuracy, epochs=10, \text{batch\_size=32}, verbose=0 \\
& \text{compute\_class\_weight} method with \text{class\_weight=balanced} \\
\hline
\end{tabular} \vspace{-2mm}
}
\end{table}
%\reviewcomment{Check the NN row: for a five-class sparse-categorical task, the output layer would normally have 5 units, not 1. If this is a typo, fix it; if not, explain the encoding and loss carefully.}
%\reviewcomment{Add a tuning-budget column or short note for every model. Reviewers need to know whether TM, XGBoost, LGBM, RF, and NN received comparable model-selection effort.}

% Table 4- Classifier performance 
\begin{table} [t!]
\caption{Model performance.}
\label{mod_perf} 
\centering
{
\setlength\tabcolsep{1.5pt}
\begin{tabular}{|c|c|c|c|c|}
\hline
Model & Precision & Recall & \text{F1-score} & Inference time \\
%\cline{2-5}
& (in \%) & (in \%) & (in \%) & (in $\mu$s) \\
\hline
TM & $97.87 \pm 0.01$ & $97.83 \pm 0.00$ & $97.83 \pm 0.00$ & $66.24 \pm 1.11$ \\
\hline
STM & $97.38 \pm 0.07$ & $97.32 \pm 0.08$ & $97.33 \pm 0.08$ & $16.89 \pm 0.29$ \\
\hline
DT & $94.67 \pm 0.20$ & $94.35 \pm 0.15$ &	$94.51 \pm 0.17$ & $0.08 \pm 0.04$ \\
\hline
RF & $95.04 \pm 0.13$ &	$95.21 \pm 0.16$ &	$95.11 \pm 0.04$ & $3.73 \pm 0.10$ \\
\hline
XGBoost & $91.75 \pm 0.16$ & $97.97 \pm 0.06$ &	$94.47 \pm 0.11$ & $1.77 \pm 0.08$ \\
\hline
LGBM & $91.74 \pm 0.25$ & $97.95 \pm 0.08$ & $94.43 \pm 0.19$ & $6.00 \pm 0.38$ \\
\hline
KNN & $93.66 \pm 0.14$ & $93.10 \pm 0.27$ &	$93.36 \pm 0.15$ & $186.44 \pm 3.19$ \\
\hline
NB & $49.91 \pm 2.70$ & $56.81 \pm 0.31$ & $44.76 \pm 2.38$ & $0.65 \pm 0.03$ \\
\hline
LR & $58.58 \pm 0.12$ & $77.10 \pm 0.31$ & $61.19 \pm 0.21$ & $0.05 \pm 0.00$ \\
\hline
NN & $82.45 \pm 0.65$ & $95.19 \pm 0.43$ & $87.29 \pm 0.74$ & $6.76 \pm 0.75$ \\
\hline
\end{tabular} \vspace{-2mm}
}
\end{table}
%\reviewcomment{This table would be stronger with macro, micro, and weighted F1, class-wise recall for minority classes, and a definition of how inference time was measured. The LaTeX header also needed adjustment because \texttt{\textbackslash SI\{\}\{\textbackslash micro\textbackslash second\}} caused pdflatex to fail here.}
%\reviewcomment{Explain the near-zero standard deviations, especially for TM recall/F1 reported as $\pm 0.00$. If values are rounded, report more decimals or confidence intervals; if folds are nearly identical, explain why.}

% Table 5- Class-wise precision, recall and F1-score for TM 
\begin{table} [t!]
\caption{Class-wise performance of the TM model.}
\label{class_wise_perf_TM} 
\centering
{
\setlength\tabcolsep{2.0pt}
\begin{tabular}{|c|c|c|c|}
\hline
Class & Precision (in \%) & Recall (in \%) & \text{F1-score} (in \%) \\
\hline
Benign & 99.0 & 100.0 & 99.0 \\
\hline
Reconnaissance & 99.0 & 97.0 & 98.0 \\
\hline
Initial access & 100.0 & 99.0 &	100.0 \\
\hline
Lateral movement & 94.0 & 97.0 & 96.0 \\
\hline
Exfiltration & 98.0 & 95.0 & 96.0 \\
\hline
\end{tabular} \vspace{-6mm}
}
\end{table}

\begin{comment}
% Fig.5: Training and Testing accuracy of TM model
\begin{figure}[t!]
\centering
\includegraphics[width=0.9\columnwidth,height=3.8cm,keepaspectratio]{Figure/tm_accuracy_vs_epochs_last_fold_TM.pdf} 
\caption{Training and testing accuracy of the TM model.}
\label{train_test_acc_TM} \vspace{-3mm}
\end{figure}
%\reviewcomment{A learning curve from the last fold is useful but insufficient for the overfitting claim. Add fold-averaged curves or report train/test gaps across all folds.}
\end{comment}

% Fig.5: Confusion matrix of TM model
\begin{figure}[t!]
\centering
\includegraphics[width=0.95\columnwidth,height=4.95cm,keepaspectratio]{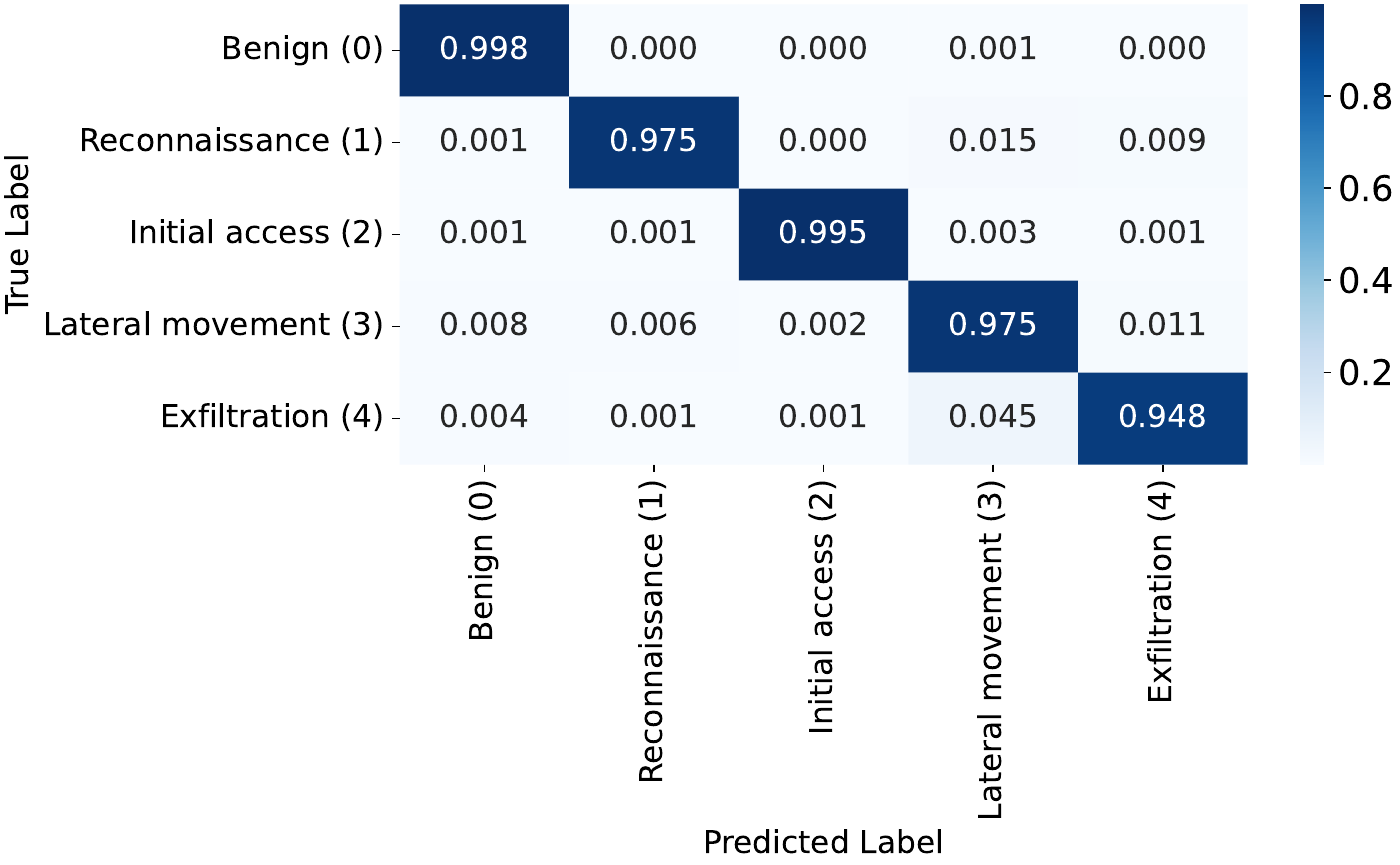} 
\caption{TM confusion matrix.}
\label{cm_TM} \vspace{-3mm}
\end{figure}
%\reviewcomment{Make sure the confusion matrix is readable in the final PDF. Consider adding a small table of per-class recall/F1 because reviewers may not be able to extract exact values from the figure.}

% Fig.6: Feature contribution of a benign sample 14596
\begin{figure*}[t!]
\centering
\includegraphics[width=\textwidth,height=5.0cm,keepaspectratio]{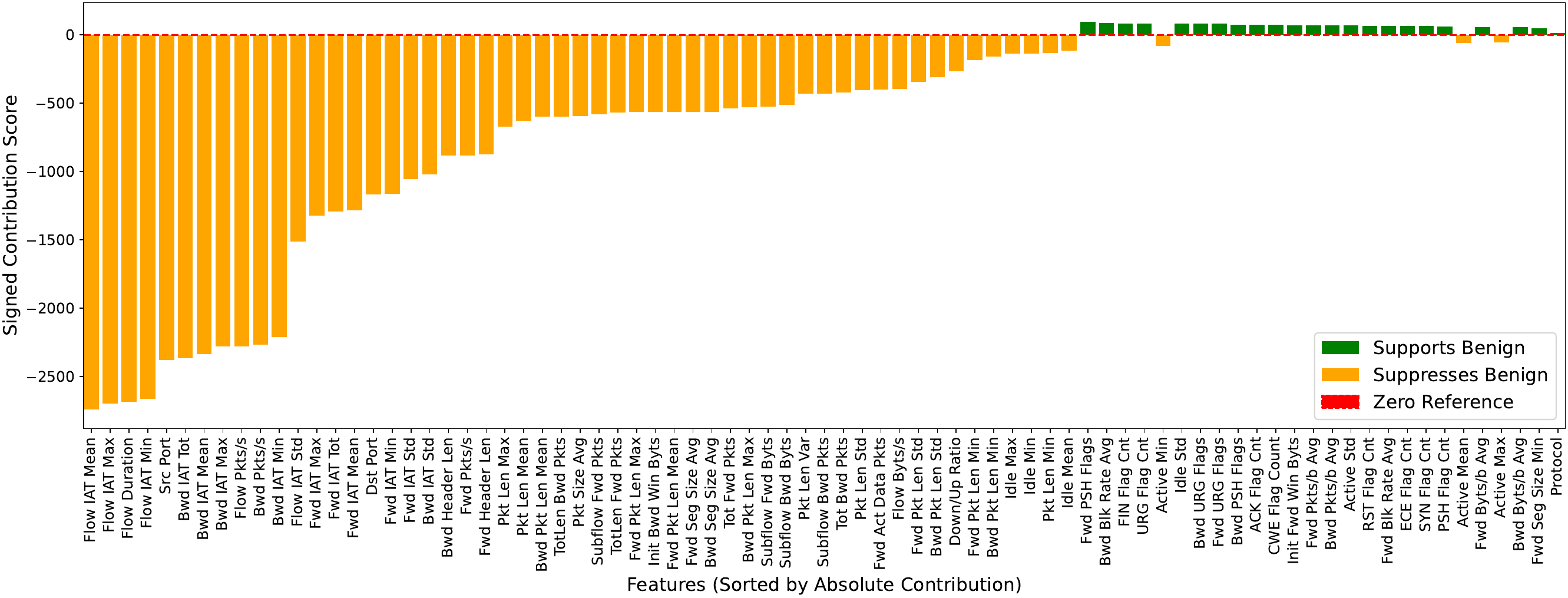} 
\caption{Feature-level contribution of a Benign sample.}
\label{feat_contri_benign} \vspace{-3mm}
\end{figure*}

% Sample number 14596
% Fig.7: Class-wise votes of a benign sample 14596
\begin{figure}[t!]
\centering
\includegraphics[width=0.95\columnwidth,height=3.5cm,keepaspectratio]{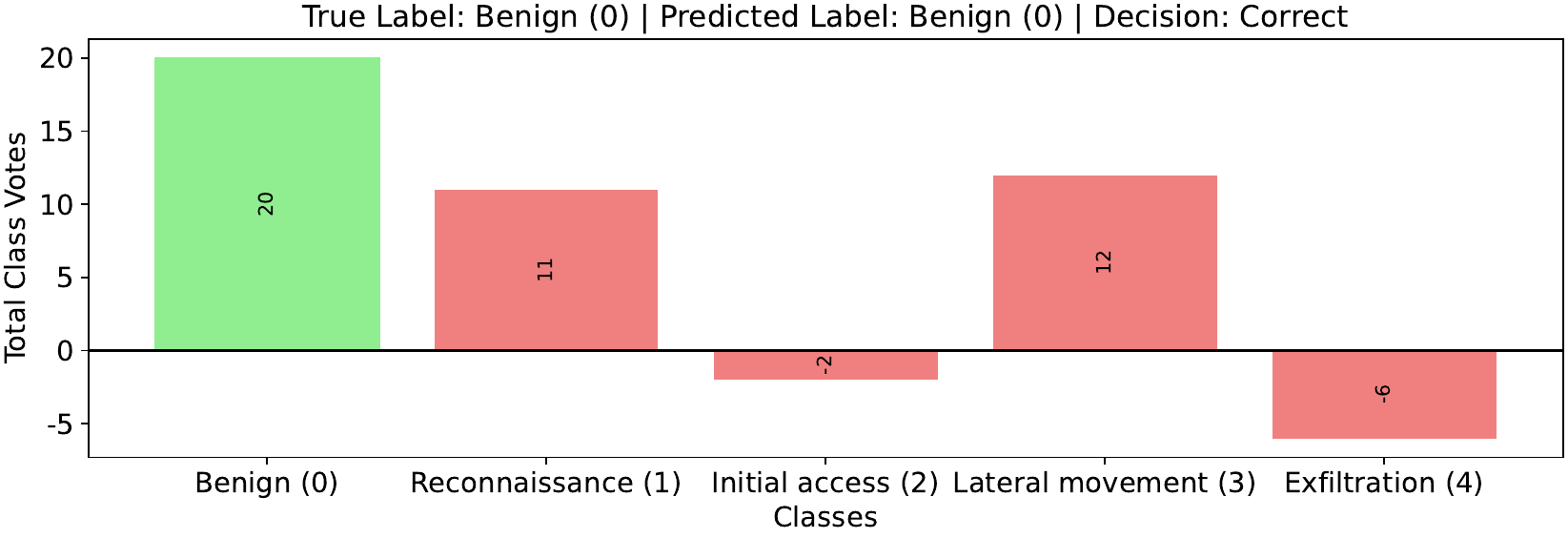} 
\caption{Class-wise votes of the same Benign sample.}
\label{cls_vote_benign} \vspace{-3mm}
\end{figure}

% Fig.8: Clause activation heatmap of a benign sample 14596
\begin{figure}[t]
\centering
\includegraphics[width=0.95\columnwidth,height=3.75cm,keepaspectratio]{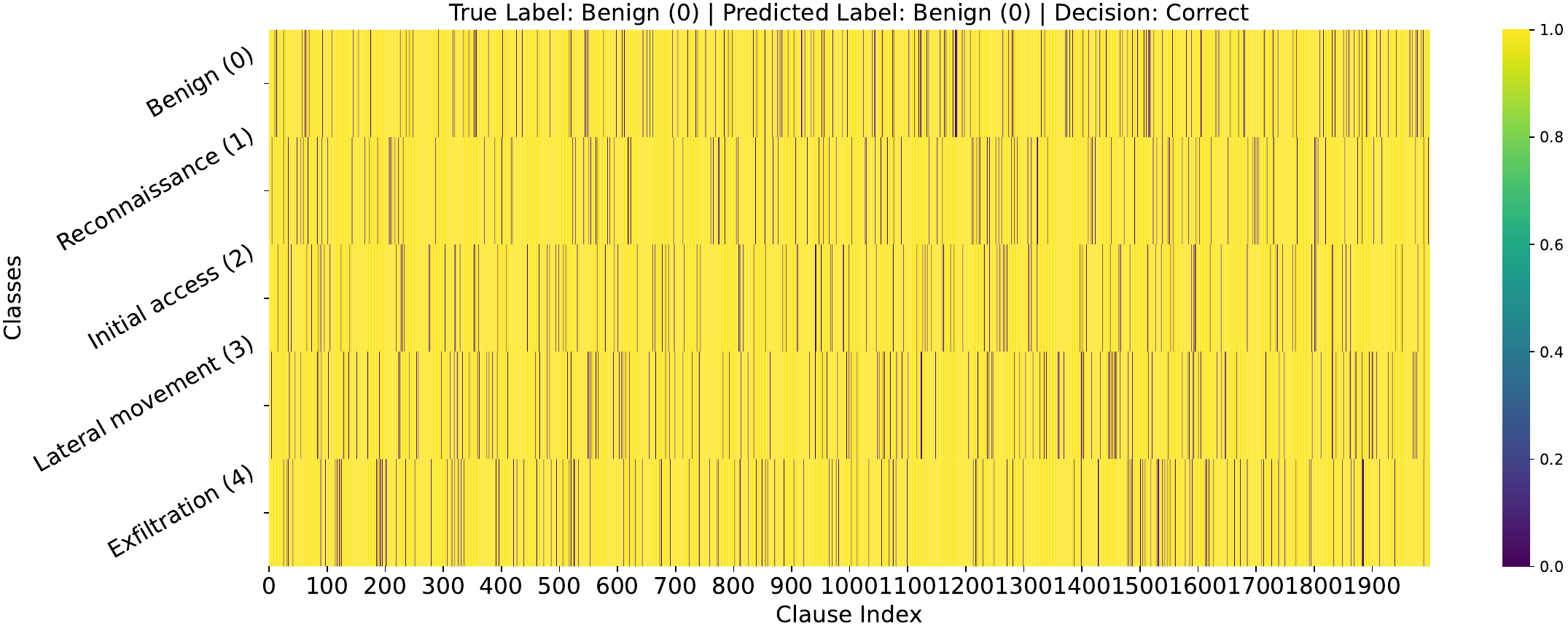} 
\caption{Clause activation heatmap of the same Benign sample.}
\label{cls_heatmap_benign} \vspace{-3mm}
\end{figure}

% Explanation for Table 4 
Table~\ref{mod_perf} indicates that most classifiers achieve consistently high mean values of precision, recall, and F1-score with minimal variance, except for Naive Bayes and Logistic Regression, which show comparatively lower performance. These results highlight the strong capability of the models to distinguish between benign traffic and various attack classes. Among all models, the TM model achieves the highest F1-score of 97.83\%, demonstrating superior performance. However, this comes with a slightly higher inference time of 66.24~$\mu$s. Although XGBoost and LGBM achieve slightly higher recall (97.97\% and 97.95\%, respectively), their lower precision indicates a higher false-positive rate. In contrast, TM provides a better precision-recall balance, resulting in the highest F1-score. Note that standard deviations are rounded to two decimal places; consequently, very small variations may appear as 0.00. In contrast, Logistic Regression offers the fastest inference time of 0.05~$\mu$s, but its F1-score is limited to 61\%. The Sparse TM (STM) achieves a reduced inference time (16.89~$\mu$s) compared to the TM model (66.24~$\mu$s), as expected, while maintaining competitive performance.
%\reviewcomment{Discuss precision-recall trade-offs, not only the highest F1. XGBoost and LGBM have higher recall but lower precision; in healthcare IDS, the preferred operating point depends on the cost of false positives versus missed attacks.}

% Explanation for Table 5
Table~\ref{class_wise_perf_TM} indicates that the TM model achieves consistently high recall (95 to 100\%) and F1-scores (96 to 100\%) across all classes, showing balanced performance across all classes and effective detection of benign traffic and attack classes.

% Explanation for Fig.5
Next, Fig.~\ref{cm_TM} presents the confusion matrix of the TM model, highlighting strong diagonal dominance and accurate classification across all five attack classes. The model achieves very high true positive rates for benign and initial access traffic. Similarly, reconnaissance and lateral movement are classified effectively, with only minor confusion between related attacks. The exfiltration class exhibits limited confusion with lateral movement, with 4.5\% of exfiltration samples misclassified as lateral movement, likely due to similarities between these attack phases. Overall, the TM model shows reliable detection of various phases of cyberattacks.
%\reviewcomment{Quantify the exfiltration-to-lateral-movement confusion. A short sentence with exact counts or percentages would make the error analysis much more convincing.}

% Interpretability of TM 
\subsubsection{TM Interpretability} 
\begin{figure}[t!]
\centering
\includegraphics[width=0.95\columnwidth,height=3.5cm,keepaspectratio]{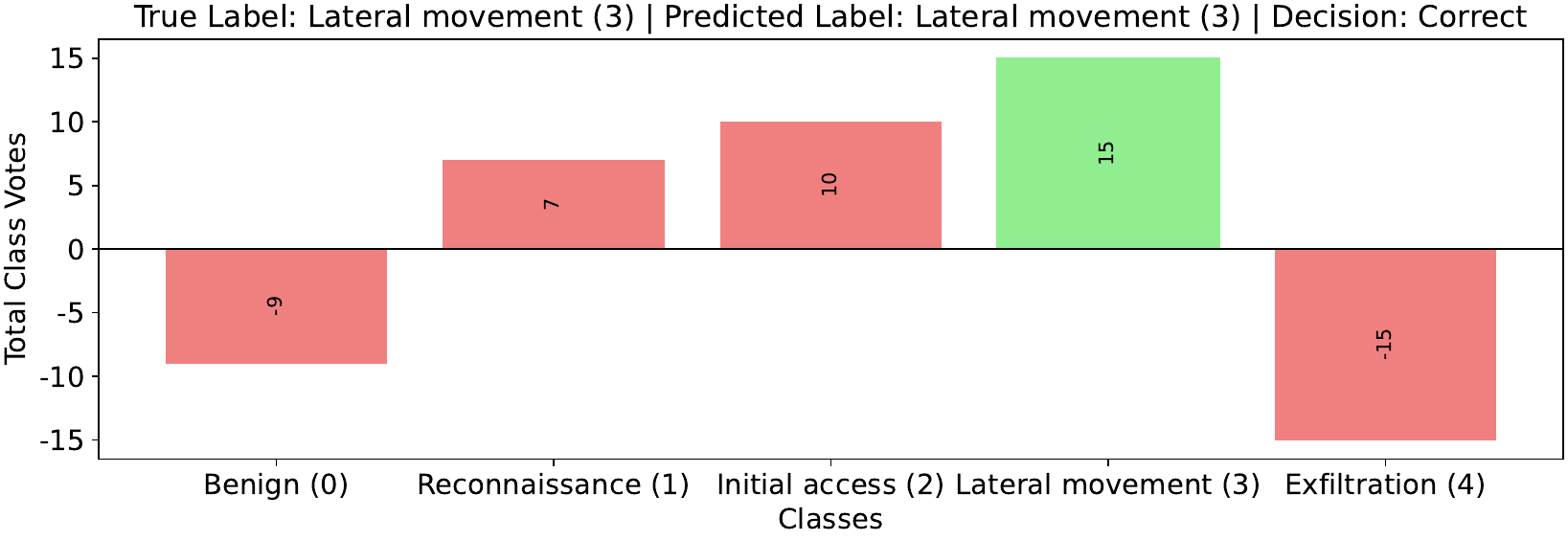} 
\caption{Class-wise votes of the Lateral movement sample.}
\label{cls_vote_lat_mov} \vspace{-3mm}
\end{figure}

\begin{figure}[t!]
\centering
\includegraphics[width=0.95\columnwidth,height=3.5cm,keepaspectratio]{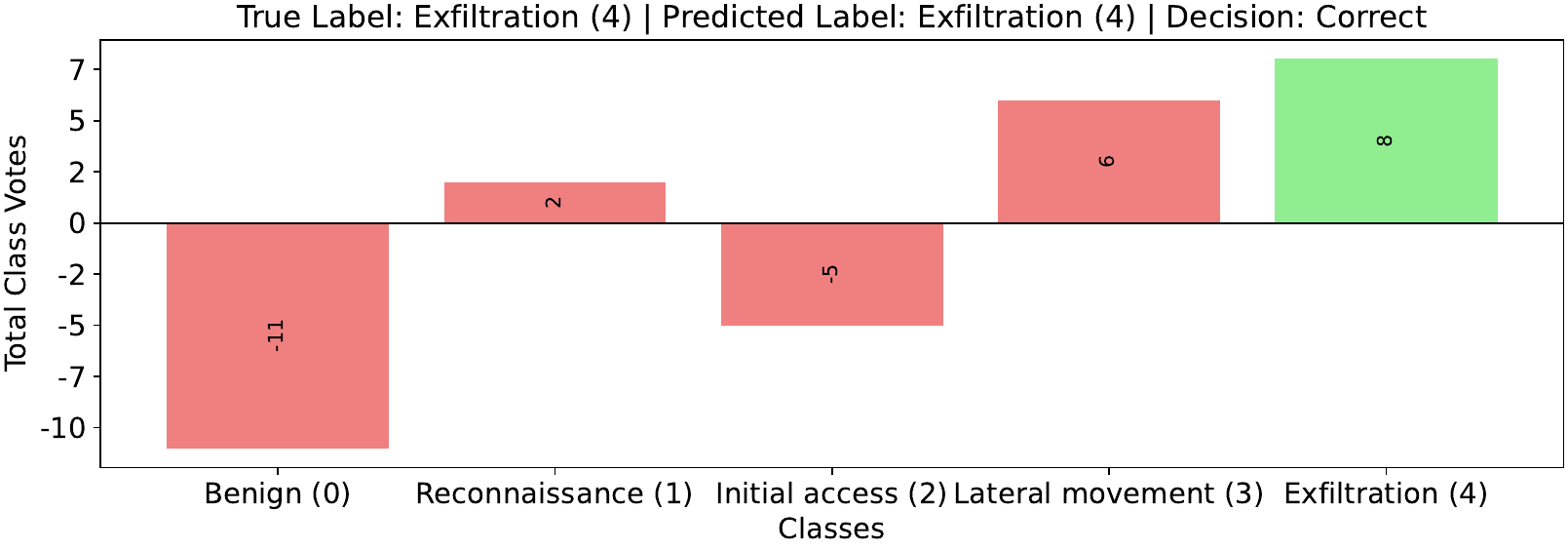} 
\caption{Class-wise votes of the Exfiltration sample.}
\label{cls_vote_exf} \vspace{-3mm}
\end{figure}

\begin{figure}[t!]
\centering
\includegraphics[width=0.95\columnwidth,height=3.5cm,keepaspectratio]{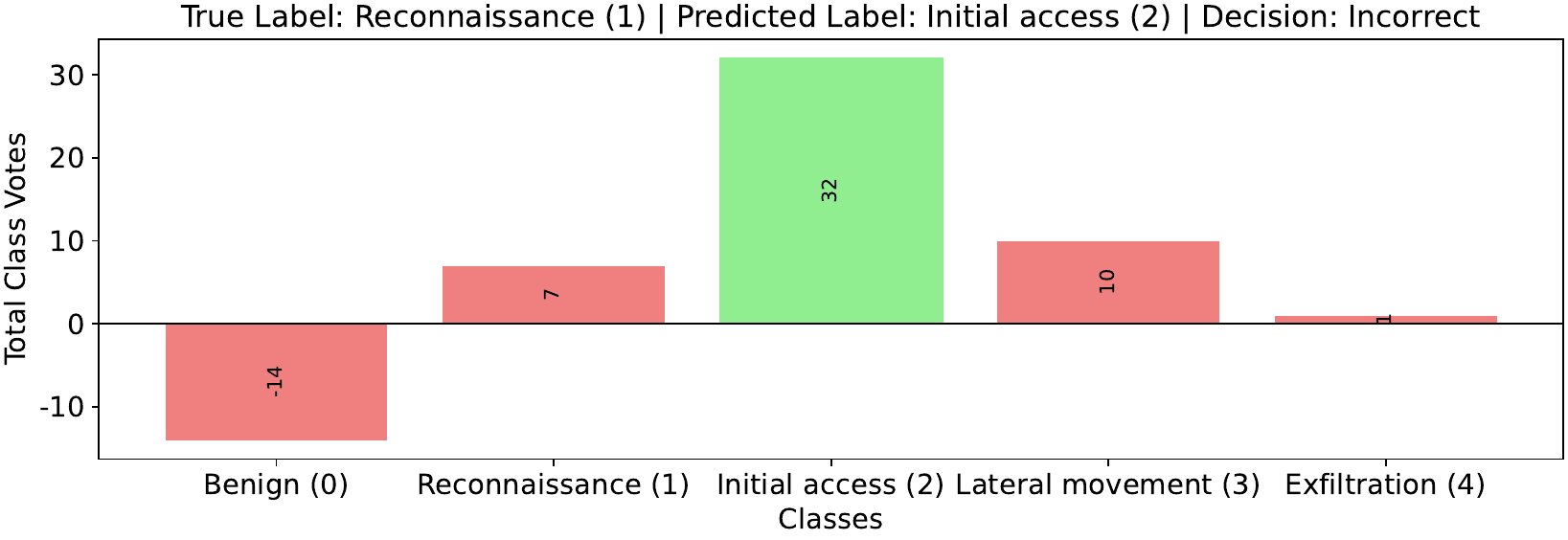} 
\caption{Class-wise votes of the Reconnaissance sample.}
\label{cls_vote_rec_ini} \vspace{-3mm}
\end{figure}

To demonstrate the interpretability of the TM model’s decisions, Figs.~\ref{feat_contri_benign}, \ref{cls_vote_benign}, and \ref{cls_heatmap_benign} illustrate the feature-level contributions, class-wise vote scores, and clause activation heatmap for a benign test sample, respectively. Fig.~\ref{feat_contri_benign} shows that most features contribute negatively (orange), suppressing non-benign patterns, while a smaller subset provides positive support (green) for the benign class. This indicates that the TM model achieves correct classification by effectively suppressing conflicting patterns. Fig.~\ref{cls_vote_benign} shows that the benign class attains the highest class vote (20), indicating normal traffic behaviour. Fig.~\ref{cls_heatmap_benign} shows that each cell represents the clause activation status for a given sample, where yellow (1) denotes active clauses and dark purple (0) denotes inactive clauses. The benign class shows a higher number of activated positive clauses, contributing to a larger class vote, while other classes show fewer clause activations. This difference in clause activity results in the highest class vote for the benign class, leading to correct classification. Similarly, Fig.~\ref{cls_vote_lat_mov} and Fig.~\ref{cls_vote_exf} illustrate that the lateral movement and exfiltration classes attain the highest votes of 15 and 8, respectively, compared to other classes, indicating correct malicious traffic detection. On the contrary, Fig.~\ref{cls_vote_rec_ini} shows a misclassification example where a reconnaissance instance is predicted as initial access. The initial access class obtains the highest class votes (32), compared to 7 votes for the correct reconnaissance class. The error likely stems from similarities in the traffic characteristics of the two attack phases, causing the TM to favor the initial access class.
%\reviewcomment{The interpretability section would be stronger if it showed what a domain user learns. Add 2-3 concrete clauses or features and explain why they make cybersecurity sense.}
%\reviewcomment{Include at least one hard or misclassified sample. Interpretability is more persuasive when it explains both a correct case and a failure mode, especially where exfiltration is confused with lateral movement.}

% Sample number 599949
% Fig.9: Class-wise votes of a Lateral movement sample 599949 

% Sample number 786919
% Fig.10: Class-wise votes of an Exfiltration sample 786919

% Sample number 166786
% Fig.11: Class-wise votes of a Reconnaissance sample 166786

\subsection{On-Device Analysis} % ok
\label{edg_dep}
Table~\ref{edge_perf} shows that the proposed TM and STM provide a balanced trade-off between performance and resource consumption on the Raspberry Pi. While both require higher memory usage than DT, NB, and LR, they maintain moderate CPU usage (around 25\%) and smaller model sizes than RF and KNN. In contrast, XGBoost and LGBM achieve fast inference at the cost of very high CPU usage (around 99\%). The STM further improves computational efficiency over TM, supporting real-time edge deployment. The CPU and memory usage are measured using \texttt{psutil} during inference. Despite its smaller model size, STM exhibits higher runtime memory usage, likely due to sparse-representation overhead.
%\reviewcomment{The edge claim needs measurement context: number of runs, batch size, monitoring interval, warm-up, whether CPU is one core or total, and whether feature extraction cost is included.}
%\reviewcomment{Report throughput in flows per second and end-to-end latency if claiming real-time IDS. Per-sample model inference alone does not establish real-time operation.}

% Table 6- Edge performance 
\begin{table} [t!]
\caption{Edge performance.}
\label{edge_perf} 
\centering
{
\setlength\tabcolsep{4.0pt}
\begin{tabular}{|c|c|c|c|c|}
\hline
Model & Inference time & Memory usage & CPU usage & Model size \\
& (in $\mu$s) & (in KB) & (in \%) & (in KB) \\
\hline
TM & 230.44 & 161920 & 25.5 & 11075.09 \\
\hline
STM & 222.76 & 273360 & 25.9 & 6028.08  \\
\hline
DT & 0.26 & 928 & 30.8 & 915.25 \\
\hline
RF & 17.77 & 128 & 25.5 & 127051.40 \\
\hline
XGBoost & 8.02 & 752 & 98.9 & 5920.76 \\
\hline
LGBM & 58.81 & 480 & 99.2 & 3455.24 \\
\hline
KNN & 3615.08 &	2256 & 99.2 & 277093.31 \\
\hline
NB & 4.57 & 1 & 28.0 & 7.01 \\
\hline
LR & 1.05 & 12208 & 91.5 & 4.23 \\
\hline
NN & 58.40 & 43856 & 37.9 & 110.84 \\
\hline
\end{tabular} %\vspace{-1mm}
}
\end{table}
%\reviewcomment{Check the memory and CPU figures for plausibility and explain how they were measured. STM has a smaller model size than TM but higher memory usage, which may be true, but needs interpretation.}

\subsection{State-of-the-Art Comparison} % ok
\label{comp_sota}
Table~\ref{comp_soa} shows that the proposed TM-based IDS achieves comparable performance to the reported ML classifier on the same dataset while providing interpretable rule-based classification. In addition, the proposed TM models demonstrate feasibility for edge deployment, making it a practical and transparent solution for IoMT cyberattack detection.

%\reviewcomment{The comparison should be framed narrowly unless more methods are included. At present, it is a same-dataset comparison against one reported ML classifier, and the F1 margin is only 0.05 percentage points.}
%\reviewcomment{The stronger claim here is not a clear performance win; it is comparable performance plus interpretability and edge feasibility. Reframe the paragraph around that contribution.}

% Table 7- Comparison with SOTA methods
\begin{table} [t!]
\caption{Comparison with the state-of-the-art methods.}
\label{comp_soa} 
\centering
{
\setlength\tabcolsep{2.0pt}
\begin{tabular}{|c|c|c|c|c|c|}
\hline
& Model & Precision & Recall & \text{F1-score} & Interpretability \\
\hline
Paper~\cite{almobaideen2025medsec} & ML Classifier & 97.73\% & 97.83\% & 97.78\% & No \\
\hline
Proposed & Sparse TM & 97.38\% & 97.32\% & 97.33\% & Yes \\
\hline
Proposed & TM & 97.87\% & 97.83\% & 97.83\% & Yes \\
\hline
\end{tabular} \vspace{-3mm}
}
\end{table}
%\reviewcomment{Add the evaluation protocol used by the cited MedSec-25 baseline if available. The same dataset is not enough for a fair comparison if splits, preprocessing, or metric averaging differ.}

\section{Conclusions and Future Work} % ok
\label{con_fut}
This work addresses the challenge of detecting various cyberattack phases in IoMT networks by proposing a TM-based IDS. The model employs an interpretable rule-driven learning framework and is trained and evaluated on the MedSec-25 dataset containing realistic attack phases. Experimental results demonstrate competitive performance with the evaluated ML classifiers, while providing transparent and interpretable decision-making. The model is further deployed on a Raspberry Pi, demonstrating the feasibility of low-latency edge inference. These characteristics make it a promising solution for IoMT healthcare services, requiring trust, reliability, and timely decision-making. This study is limited to a single public dataset, offline traffic analysis, limited external validation, and interpretability demonstrated through selected examples rather than a formal user-based evaluation, which may limit the generalizability of the reported performance to other IoMT environments. Future work will extend the evaluation to self-developed datasets with broader attack variability, investigate user-centered assessment of interpretability, and analyze the sensitivity of TM performance to key hyperparameters such as $s$ and $T$.

%\reviewcomment{Add a limitations sentence before future work covering the single public dataset, offline traffic, limited external validation, and interpretability demonstrated on selected examples rather than formally evaluated with users.}
%\reviewcomment{Tone down the conclusion until the protocol is fully specified. Replace broad wording such as "performs better than ML classifiers and existing state-of-the-art methods" with a claim tied exactly to the reported table.}

%\begin{comment}
\section*{Acknowledgement}
This publication has emanated from the research project SecureIoTM: Ultra-low-energy IoT Intrusion Detection Systems using Logic-based Tsetlin Machines, under Grant Number 342167, funded by the Research Council of Norway.
%\end{comment}

% Reference section
%\reviewcomment{All cited keys are present, but the bibliography contains several uncited entries. Also, verify high-risk web or current references before submission.}
\bibliographystyle{IEEEtran}
\bibliography{references}

\end{document}